\documentclass[10pt,twocolumn,twoside]{IEEEtran}

\usepackage{graphicx}

\usepackage{cite}

\newcommand {\col}[1]{{\bf{#1}}}
\newcommand {\mat}[1]{{\bf{#1}}}

\newcommand {\p}{p}
\newcommand {\I}{P}
\newcommand {\gp}{r}
\newcommand {\GP}{R}

\newcommand {\dt}{\Delta t}
\newcommand {\df}{\Delta f}
\newcommand {\dM}{K}

\newcommand {\s}{s}

\begin{document}

\title{Real-Time Calibration of the Murchison Widefield Array}

\author{D. A. Mitchell, L. J. Greenhill, R. B. Wayth, R. J. Sault, C. J. Lonsdale, \\
        R. J. Cappallo, M. F. Morales, and S. M. Ord}

\markboth{IEEE JSTSP,
          Special Issue on Signal Proc. for Astronomical and Space Research Applications}
         {Mitchell \MakeLowercase{\textit{et al.}}: Real-Time Calibrator Measurements}

\maketitle

%----------------------------------------------------------------------------------------------------------------------%

\begin{abstract}

The interferometric technique known as peeling addresses many of the challenges faced when observing with low-frequency
radio arrays, and is a promising tool for the associated calibration systems. We investigate a real-time peeling
implementation for next-generation radio interferometers such as the Murchison Widefield Array (MWA). The MWA is being
built in Australia and will observe the radio sky between 80 and 300 MHz. The data rate produced by the correlator is
just over 19 GB/s (a few Peta-Bytes/day). It is impractical to store data generated at this rate, and software is
currently being developed to calibrate and form images in real time. The software will run on-site on a high-throughput
real-time computing cluster at several tera-flops, and a complete cycle of calibration and imaging will be completed
every 8 seconds. Various properties of the implementation are investigated using simulated data. The algorithm is seen
to work in the presence of strong galactic emission and with various ionospheric conditions. It is also shown to scale
well as the number of antennas increases, which is essential for many upcoming instruments. Lessons from MWA pipeline
development and processing of simulated data may be applied to future low-frequency fixed dipole arrays.

\end{abstract}

%\begin{IEEEkeywords}
%Antenna arrays, array signal processing, calibration, dipole arrays, radio astronomy, radio interferometry.
%\end{IEEEkeywords}

%----------------------------------------------------------------------------------------------------------------------%

\section{Introduction}
\label{Introduction}

\begin{figure}[ht]
  \centering
  \includegraphics[width=7.8cm]{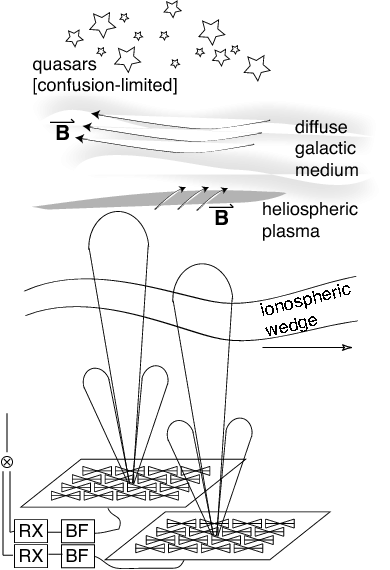}

  \caption{Schematic diagram of an MWA interferometer baseline and sky. Each MWA receiving element is a phased tile
           comprising a $4\times4$ grid of crossed dipoles ({\it vertical planar bowtie structures}). The main response
           lobes are steered electronically by beam-formers ({\it BF}) to establish the instrument field of view. BF
           output signals are sampled at baseband and filtered digitally in receiver electronics ({\it RX}) and
           correlated to provide cross-power spectra. In addition to compact objects (e.g., quasars and pulsars), the
           MWA sky includes foreground emission from the galaxy, which is chiefly synchrotron emission that is linearly
           polarized and traces turbulent and ordered magnetic fields in the interstellar medium ($\vec{B}$). Magnetized
           plasma from solar coronal mass ejections or other activity move outward through the heliosphere. This medium
           and the Earth ionosphere induce primarily refractive shifts in source positions (higher order distortions in
           the observing passband are small due to the limited extent of the array) and time-variable Faraday rotation
           of polarization along different lines of sight. The calibration scheme described here enables
           polarization-sensitive imaging of the diffuse galactic and compact source populations and assembly of a sky
           model. Knowledge of the compact background sources is ``confusion limited'' (i.e., constrained by
           multiplicity of sources within one instrument resolution element and contamination of each image pixel by
           sidelobes of distant sources). Confusion increases with image sensitivity and the size of a resolution
           element..}

  \label{plot:schematic}
\end{figure}

The Murchison Widefield Array (MWA) is an 80-300 MHz synthesis array that is being built in Western Australia, with
construction to be completed in 2010. The shire of Murchison has a quiet radio environment, making it an excellent site
for this and other radio facilities \cite{Bowman2007a}. Each of the 512 antennas will be a $4\times4$ tile of dipoles.
An analogue beamformer at each antenna combines the signals from the 16 dipoles, producing an electronically steerable
primary beam with a width of approximately 25$^\circ$ at 150 MHz. When the signals from all antennas are combined, the
array will have a synthesized beam with a width of approximately 4.5$^\prime$ at 150 MHz. The main science goals of the
MWA are the detection of redshifted 21cm emission from the Epoch of Reionization (EoR) \cite{Bowman2006}, transient
detection (for example \cite{Bhat2007}), and remote heliospheric sensing \cite{Salah2005}. A schematic of two MWA
antenna tiles is shown in Fig. \ref{plot:schematic}.

To make a map of the sky using radio interferometry, one typically builds up an estimate of the 2D Fourier transform of
the sky, then applies a Fourier transform to obtain the image. This is known as synthesis imaging, and a good overview
of the subject is given in \cite{Thompson-etal.2001}. The measured data that are used to build up the Fourier
interference pattern are spatial cross-correlations -- or visibilities -- that are obtained by correlating voltage
streams collected by many pairs of antennas. The visibilities lie in the uvw coordinate frame, where $w$ is the
component of the antenna separation vector (or \emph{baseline} vector) in the direction of the field center (in units of
wavelengths), and $u$ and $v$ are orthogonal coordinates in the plane normal to $w$ (aligned with the corresponding
image coordinate axes, $l$ and $m$). This situation is not naturally a 2D Fourier transform. For small images, $w$ is
multiplied by a term that is approximately zero and the 2D nature holds. For large fields this is not the case, but the
problem can still be reduced to 2D transforms (a good overview is given in \cite{Cornwell2005}).\footnote{The MWA will
produce \emph{snapshot} images, which means each antenna pair contributes a single visibility to each image. The MWA
visibilities will be approximately coplanar, so a 2D Fourier relationship will hold even for wide-field images.} Post
processing typically involves calibrating the visibilities, gridding them onto the uv-plane to form a regularly sampled
interference pattern, and then applying a 2D FFT to form an image. Techniques such as self-calibration can then be used
in an attempt to improve the calibration by iterating back and forth between the visibilities and the image.

For a number of reasons, MWA visibilities cannot be processed in this way. Many of these effects are common to all low
frequency arrays, and are described in detail in \cite{Erickson1999} and \cite{Thompson-etal.2001}. Each antenna has a
different direction-dependent response over the field of view, which cannot simply be divided out. These response
patterns may also change significantly over the course of an observation. Furthermore, the ionosphere causes
direction-dependent phase shifts that effectively change the position and polarization state of sources during an
observation. These effects mean that we cannot make a fully calibrated interference pattern in the standard way. What we
can do is use the measured visibilities to iteratively fit ionospheric phase shifts and antenna gains towards many
bright catalogue sources, and store these fits to aid deconvolution and resampling processes after the images have been
made. These measurements are the focus of this paper, but before they are discussed some of these challenges will be
reviewed more closely.

Unlike the radio sky at higher frequencies, which appears sparsely populated at the sensitivity levels of modern
instruments, the sky to be observed by the MWA is full of sources. The high density of sources and large angular
resolution of the array will result in images that are confusion limited, with significant flux coming from background
galaxies in each synthesized beam, as well as from the sidelobes of other sources in the primary beam of each antenna.
The sky is also quite complex. There is significant emission on many angular size scales, from compact extragalactic
sources and pulsars to diffuse galactic synchrotron radiation \cite{Erickson1999}. The latter experiences Faraday
rotation and depolarization in the ionized interstellar medium, and as a result exhibits a strong linearly polarized
component that is highly dependent on position and frequency \cite{Wieringa1993}. This polarized all-sky signal is at
least a few orders of magnitude brighter than the unpolarized EoR signal, and a fully-polarized calibration formalism,
such as the Hamaker-Bregman-Sault measurement equation \cite{Sault-etal.1996}, is required to reduce contamination from
the spatially structured, linearly polarized emission from the galaxy.

If the raw interferometer data are stored for offline processing, iterative calibration and deconvolution algorithms can
be used to address many of the problems described below (see, for example, \cite{Bhatnagar2008} and \cite{Nijboer2007}).
However, it is impractical to store the 19 GB s$^{-1}$ data stream coming from the MWA correlator,\footnote{The
$N_a=512$ antennas lead to $N_a(N_a-1)/2=130816$ different cross-correlation measurements. They are made in a correlator
for 4 different polarization products and up to 3072 different frequency channels every 0.5 seconds. Each visibility is
represented by 3 real bytes and 3 imaginary bytes, which leads to $\sim19$ GB/s.} and the MWA will store images. This
means that much of the calibration must take place in real time, before or during the imaging process. At the heart of
the calibration system for the MWA is the calibrator measurement loop (CML), which measures apparent angular offsets
induced by the ionosphere and the system gain toward known compact astronomical sources across the sky. These
measurements are used to fit models of the ionosphere and instrument response, and support subtraction of strong sources
that limits sidelobe contamination during calibration. As in \cite{Noordam2004} and \cite{vanderTol2007}, measuring and
subtracting the contribution of each source is carried out sequentially, so that the stronger sources are removed before
measurements of weaker sources are made. In this paper we do not consider multivariate fits of parameters for all of the
calibrator sources simultaneously, as described in \cite{vanderTol2007}, but we will discuss it briefly in section
\ref{Discussion}.

Estimation of calibration parameters is greatly over constrained due to the large number of antenna pairs
($1.31\times10^5$). On the other hand, the wide-field nature of the instrument and real-time computing requirement pose
challenges, the most important of which are listed below.

\begin{enumerate}

  \item \emph{Direction-dependent gain and polarization response}. Each MWA receiving element is a $4\times4$ array of
        fixed crossed dipoles (Fig. \ref{plot:schematic}). The phased beams are steered with an analogue beamformer,
        which will typically be updated every 5 to 10 minutes to compensate for rotation of the Earth. The common
        approach of assuming that the polarized receptors are orthogonal over the field of view with a small amount of
        direction-dependent leakage cannot be used. The direction-dependent instrumental polarization of the antenna
        beams will be significant, and it will be measured along with the direction-dependent gain using many calibrator
        sources spread over the entire sky. These measurements will be repeated as the field of interest moves across
        the antenna beams.

  \item \emph{Confusion}. Since the MWA's primary beams cover such a large section of the sky, each field mapped by the
        MWA will contain hundreds of relatively bright sources. To calibrate the array, we require accurate flux density
        measurements of known sources. Such measurements can be corrupted by faint sources within the synthesized beam
        of the array (``confusion'') as well as the sidelobes of brighter sources outside the synthesized beam
        (``sidelobe contamination'').

        Confusion and sidelobe contamination can arise from both compact and large-scale sources such as extragalactic
        radio galaxies and galactic synchrotron emission respectively. The Galactic synchrotron, in particular, has a
        polarized component and structure on many spatial scales. Since the interferometer baselines will respond to
        large-scale structure differently depending on their length and orientation, calibration of data that includes
        bright resolved sources such as the Sun or Galactic plane must be performed carefully.

  \item \emph{Ionosphere}. The 3-dimensional ionosphere can significantly perturb the waves coming from celestial
        sources. The maximum antenna separation of the MWA is short enough that, for a given source and during normal
        ionospheric conditions, all of the antennas have approximately the same line of sight through the ionosphere.
        This assumption will be used throughout. Under these conditions there will be no defocusing and the ionosphere
        can be described by a two-dimensional phase screen that makes sources appear to move away from their true
        positions. (Faraday rotation of incident polarization is a second ramification, but this will be considered in a
        future paper. We limit consideration here to ionospheric calibration using unpolarized sources, which cannot be
        used to calibrate polarization position angles \cite{Sault-etal.1996}, \cite{Hamaker2000a}.)

  \item \emph{Real-time data reduction}. The real-time nature of the MWA means that compute-intensive processes need to
        avoided where possible. Unfortunately, this means that many of the promising techniques currently being
        investigated, such as iterative self-calibration and deconvolution algorithms \cite{Bhatnagar2008}, and some
        wide-field imaging algorithms \cite{Cornwell2005}, cannot currently be implemented in real time. They either
        cannot be used at all or need to be approximated.

\end{enumerate}

Calibration occurs in a back-end known as the real-time system (RTS), which consists of a visibility integrator (time
and frequency), the CML, and an imaging pipeline. These tasks run sequentially, and as mentioned later the processing
load is split over frequency. The imaging pipeline incorporates gridding, imaging FFTs, correction for ionospheric and
wide-field distortion of the sky, Stokes conversion of images,\footnote{Stokes parameters describe the polarization
state of a signal as an unpolarized component, I, two linearly polarized components, Q and U, and a circularly polarized
component, V. They are used extensively in radio astronomy, see, for example, \cite{BornWolf1999} and
\cite{Thompson-etal.2001}.} and astronomical coordinate conversion. MWA primary science drivers require that the time
and frequency resolution are sufficient to ensure that stationary signals from sources throughout the antenna field of
view are coherent for the highest frequencies and longest antenna separations, where interference fringe phases vary
most rapidly with time.

Most of the key elements of the real-time calibration system have been coded and are regularly tested with simulated
data, as described in section \ref{Examples}. To date, the tests have focused on unpolarized cosmic signals, but the
response of the instrument polarizes the signals during reception, and the processing employs the fully-polarized
description discussed in the ensuing chapters. Apart from tolerance testing -- to determine optimal bandwidths, number
of calibrators, etc. -- and algorithm development for real-time operation, the main piece of outstanding work is the
incorporation of polarized calibrators into the system to fully constrain primary beam models and the Faraday rotation
component of the ionosphere.

While the discussion and examples given below focus on the MWA, the techniques are applicable to other low-frequency
array projects, such as the SKA Molonglo Prototype (SKAMP), the Long Wavelength Array (LWA), and the Low Frequency Array
(LOFAR). That said, the instantaneous synthesized beam of the MWA does make it particularly well suited to this type of
processing. We will not go into specific details on how the techniques can be optimized, which at any rate will be
different for the different arrays, and refer to papers such as \cite{Lonsdale2004} for an overview of the power of
large-$N$ radio arrays.

After outlining the assumptions and mathematical model in the next section, the steps in the CML are discussed in more
detail in section \ref{The Calibrator Measurement Loop}, followed by a discussion on algorithm convergence and
performance. We then finish with an analysis of peeling simulated MWA data in section \ref{Examples}.

%----------------------------------------------------------------------------------------------------------------------%

\section{The Visibilities}
\label{The Visibilities}

In \cite{Hamaker2000a}, Hamaker presents a matrix version of self-calibration that leads to a straightforward procedure
for estimating each antennas polarized response to a calibrator source. In this and the next section, the mathematical
formalism described in \cite{Hamaker2000a} is used to build up the planned implementation for the MWA. As in
\cite{Hamaker2000a}, bold uppercase variables will represent matrices, bold lowercase variables will represent column
vectors and a dagger ($\dagger$) will denote a Hermitian transpose. The input to the CML is a new set of visibilities,
measured every $\dt$ seconds and averaged into $M$ frequency channels of width $\df$ Hz. For the MWA, the data from the
correlator are averaged over $\dt=8$ seconds and $\df=40$ kHz, with $M=768$, and then sent straight to the CML. The
cadence time is set to 8 seconds in order to oversample the time-varying ionosphere, which fluctuates on timescales
shorter than a minute.

Suppose that the visibilities can be approximated by the superposition of $N_c$ unresolved calibrator sources that
suffer negligible smearing over $\df$ and $\dt$,\footnote{Following \cite{Thompson-etal.2001}, the fractional bandwidth
at 140MHz of $\sim0.03$\% causes less than a percent of decorrelation on the longest baselines for sources at the edge
of the map. The effect in the image is a broadening of those sources by less than 0.01\%. The integration time causes a
few 10s of percent of decorrelation on the longest baselines, and these will be processed at a faster cadence (2
seconds) to reduce this decorrelation down to the percent level. One should note, however, that the majority of
baselines are very short (baseline density goes as the reciprocal of baseline length squared outside a densely-packed
core), and most baselines suffer far less decorrelation.} and some additive noise from various sources including thermal
system noise, confusion from sources such as background radio galaxies, and sidelobes of extended emission and weak
point sources. Consider the contribution of one of these calibrator sources, $c$, to the visibility measured by antennas
$j$ and $k$ in the band centered at $f$ Hz. Let the column vector, $\col{\gp}_{j,c,f}$, contain the response of the two
orthogonal polarized components of receiving system $j$ (in instrumental polarization coordinates). It is equal to the
product of a $2\times2$ Jones matrix, $\mat{J}_{j,c,f}$, which contains the complex voltage gain of each polarized
receptor to each polarized component of the calibrator signal (including all instrumental effects), and the incident
$2\times1$ signal vector (in sky polarization coordinates), $\col{\p}_{c,f}$. The incident radiation can be described by
a $2\times2$ covariance matrix, which contains the flux density of the four polarization products,
$\col{\I}_{c,f}^{}=\left<\col{\p}_{c,f}^{}\col{\p}_{c,f}^\dagger\right>$, where angle brackets denote the expectation
value. In the absence of any ionospheric effects, and if the Jones matrices are constant over the time interval in which
the expectation values are estimated, the visibility matrix measured by baseline $jk$ is

\begin{equation}
\label{eqn: response}
\begin{array}{rcl}
\col{\GP}_{jk,c,f}
 & = & \left<\left(\mat{J}_{j,c,f}^{}\col{\p}_{c,f}^{}\right)
             \left(\col{\p}_{c,f}^\dagger\mat{J}_{k,c,f}^\dagger\right)\right>.\\
 &&\\
 & = & \mat{J}_{j,c,f}^{} \, \mat{\I}_{c,f}^{} \, \mat{J}_{k,c,f}^\dagger.\\
\end{array}
\end{equation}

Define $\col{\s}_{c}$ to be the expected position of source $c$, and
$\col{\s}_{c,f}^{\,\prime}=\col{\s}_{c}+\delta\col{\s}_{c,f}$ to be the apparent position of the refracted source, where
$\delta\col{\s}_{c,f}$ is a small error in the position estimate and a prime will be used to indicate that a variable
has been disturbed by the ionosphere. These position vectors can be expressed as phase shifts,
$\phi_{jk,c,f}^{\prime}=\phi_{jk,c,f}+\delta\phi_{jk,c,f}$, where $\phi_{jk,c,f}$ is given by the dot product of the
baseline vector and the calibrator position vector: $2\pi\col{u}_{jk,f}.\col{\s}_{c}$. The model for the ionospherically
disturbed visibility matrix of baseline $jk$ is the superposition of the contributions from each of the strong sources,

\begin{equation}
\label{eqn: vis}
\col{V}_{jk,f}^{\prime}=\col{{N}}_{jk,f}+\sum_{c=1}^{N_c}\col{\GP}_{jk,c,f}\exp\{-i\phi_{jk,c,f}^{\prime}\},
\end{equation}

\noindent where $\col{{N}}_{jk,f}$ contains the noise (thermal and confusion) in each polarization product. The negative
sign in the exponent follows the convention adopted in \cite{Thompson-etal.2001}, which is also the convention used by
software packages such as FFTW (a negative exponent is used when transforming from a real plane to an complex plane).

%----------------------------------------------------------------------------------------------------------------------%

\section{The Calibrator Measurement Loop}
\label{The Calibrator Measurement Loop}

The general approach chosen for the CML is similar to the peeling approach suggested in \cite{Noordam2004} and discussed
in \cite{vanderTol2007}. We track a few hundred strong points sources, making continual measurements of antenna primary
beams and the refractive effect of the ionosphere in each source direction. For this, we need a list of strong radio
sources with known positions and fluxes, which we get from the existing catalogues of the southern sky, such as
\cite{Wright1990} and \cite{Mauch2003}. However, some bootstrapping will be required for the fluxes, since no catalogue
covers the entire MWA frequency range. At least initially, we will only use unresolved point sources as calibrators,
since each can be described by a single Fourier component. The Molonglo Reference Catalogue (MRC) contains 7347 sources
with flux densities of or above 1 Jy at 408 MHz in the declination range $\delta$ = -85$^\circ$ to +18.5$^\circ$
\cite{Large1991}. Of these, about 90\% show no clear evidence of departure from point sources for the MRC beam of
$2^\prime.62\times2^\prime.86\sec(\delta+35^\circ.5)$.

The calibrator that contributes the most power to the visibilities is selected, which will typically be much stronger
than the superposition of the sidelobes from other sources. A model-dependent phase ramp is fit to the visibilities to
estimate the ionospheric refraction in the source direction, and then least squares estimates of the direction dependent
antenna gains are made. These estimates are used to create models of the contribution to the visibilities, which are
subtracted. This process is then repeated for the rest of the strong calibrators. After this, one will typically make
further ionospheric refraction and primary beam measurements without source subtraction.

The natural dimension over which to parallelize the processing is frequency. The imaging pipeline and much of the
calibration system can be run independently for different frequency channels, and one can think of the MWA real-time
computer as a system of essentially independent threads that each take a subset of the channels (say 10 consecutive 40
kHz channels) to process with the CML (each thread running on its own compute node).

\begin{figure*}[ht]
  \centering
  \includegraphics[width=15cm]{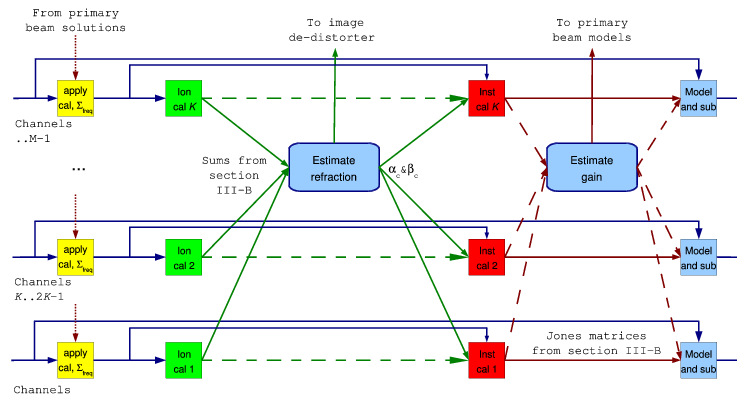}\\

  \caption{Parallelization of the CML. The processing for a single calibrator is shown, moving in sequence from left to
           right. Each horizontal chain is processed by a single compute node, with low-bandwidth intercommunication
           occurring only at a few key places. Blue lines indicate visibilities, other colors show meta-data, with
           dashed lines indicating potential data paths.}

  \label{plot:CalibrationDataflowDiagram_fig3}
\end{figure*}

To help isolate ionospheric refraction phase shifts from instrument phase shifts, the threads will be loosely coupled so
that the whole 32 MHz band can be used for the ionospheric phase measurements. While the amount of data passed between
threads is small,\footnote{The summations in (\ref{eqn: bob's solution}) need to be generated in each compute node, each
providing the information for a different observing frequency. For each source, a central node needs to gather a few 10s
of bytes from each compute node and send two fitted coefficients back.} the sources will need to be synchronized across
threads. This is shown schematically in Fig. \ref{plot:CalibrationDataflowDiagram_fig3}. There is also the potential for
pan-frequency averaging during the antenna gain measurements.

%----------------------------------------------------------------------------------------------------------------------%

%----------------------------------------------------------------------------------------------------------------------%

Let the number of consecutive frequency channels processed as a group be $\dM$, so that each of the $M/\dM$ nodes hold a
$\dM\df$ Hz sub-band. The flow of the CML in this framework is as follows.

\begin{itemize}

  \item \emph{Ranking}: Initial antenna primary beam models are used with a catalogue to rank the calibrators by
        expected received power given the current location of the source and the pointing direction of the antenna. The
        catalogue will be generated during the commissioning of the array, and the sources will be regularly surveyed to
        check for time variability. The sequential nature of the peeling and the need for synchronization across the
        full 32 MHz band mean that the different beam shapes at different frequencies need to be taken into account when
        ranking sources. The ranking will be updated at a rate given by the motion of sources through the antenna beams,
        for instance when the antenna beams are updated.

  \item \emph{Initial source subtraction}: An estimate of the calibrator summation in (\ref{eqn: vis}) is made for each
        visibility and subtracted from it (``pre-peeling''). The sum is over all of the calibrators to be peeled and the
        idea is to remove as much sidelobe power from the visibility set as possible. These estimates will usually be
        made from data measured $\dt$ seconds earlier, and provided that the ionosphere has not moved a source more than
        a synthesized beam in this time, most of the power will be removed. The strong source measurements will be less
        vulnerable to the subtraction errors of other sources, so they are made first and then peeled properly before
        the weak source measurements are made.

  \item \emph{Loop}: Each thread proceeds down the ranked calibrator list, at each step performing the following tasks.
        These tasks will be elaborated on in sections \ref{Rotate Visibilities and Sum over Frequency} through
        \ref{Source Subtraction}.

  \begin{enumerate}

    \item \emph{Rotate visibilities and sum over frequency}: Add back into the visibility set the contribution from the
          current calibrator that was subtracted during the initial source subtraction step. Set the phase center of the
          visibilities to the estimated calibrator position and average across the $\dM$ frequency channels, as shown in
          section \ref{Rotate Visibilities and Sum over Frequency}. For the MWA, some of the longer baselines will be
          stored with integration times shorter than $\dt$. These will also be averaged together at this stage. The
          averaged visibilities are saved as a new visibility set with reduced thermal noise. Averaging at this point
          also substantially reduces the number of computing operations performed by the CML.

    \item \emph{Ionospheric refraction measurements}: Each node creates intermediate sums for the quantities described
          in section \ref{Ionospheric Refraction Measurements}. These sums are gathered from each of the nodes to
          constrain the pan-frequency ionospheric refraction measurement. For isoplanatic patch sizes of $\sim4^\circ$
          and image sizes of $\sim30^\circ$, we would need at least 50 or 60 sources in the field of view to describe
          the phase variation of each patch. Optimally, we would like to oversample these variations by a factor of at
          least a few. After the fit, the parameters are broadcast back to the nodes and the phase center of each
          averaged visibility set is shifted to this new position. This is shown by green lines in Fig.
          \ref{plot:CalibrationDataflowDiagram_fig3}.

    \item \emph{Instrumental gain measurements}: Each node then performs the least squares optimization for the
          polarized voltage gain of each antenna, as described in section \ref{Instrumental Gain Measurements}. If a
          higher signal-to-noise ratio is required, the data from each sub-band can be gathered together for a full-band
          fit. This is shown by red lines in Fig. \ref{plot:CalibrationDataflowDiagram_fig3}.

    \item \emph{Source subtraction}: If the gain and ionosphere measurements pass a set of goodness-of-fit tests, they
          are used to peel the source from the full resolution visibility set, thus correcting the initial source
          subtraction that was based on old data. Otherwise the initial source subtraction is repeated. If there are
          more calibrators in the list, loop back to step 1.

  \end{enumerate}

\end{itemize}

%----------------------------------------------------------------------------------------------------------------------%

\subsection{Rotate Visibilities and Sum over Frequency}
\label{Rotate Visibilities and Sum over Frequency}

At the start of the loop for calibrator $c$, the visibilities have had all of the other calibrators peeled off --
stronger ones modeled using the current data, weaker ones modeled using less recent data. We will use
$\widehat{\col{V}}^{}_{jk,c,f_a}$ to designate the peeled visibility matrix for frequency channel $f_a$. All of the
matrices in the sub-band are rotated to be phase centered at the location of the calibrator and summed across frequency,

\begin{equation}
\label{eqn: rotated vis}
\begin{array}{rcl}
\col{V}_{jk,c,f_0}^{\prime(\dM)}
&    =    & \displaystyle\frac{1}{\dM}
            \sum_{a=1}^{\dM}\frac{\widehat{\col{V}}^{\prime}_{jk,c,f_a}}{\eta_{jk,c,f_a}}\exp\{+i\phi_{jk,c,f_a}\} \\
&&\\
& \approx & \displaystyle\col{\GP}_{jk,c,f_0}\exp\{-i\delta\phi_{jk,c,f_0}\} + \col{{N}}_{jk,c,f_0}^{\,(\dM)} ,
\end{array}
\end{equation}

\noindent where the superscript $(\dM)$ indicates an average over $\dM$ frequency channels, $\eta_{jk,c,f}$ is the
product of any undesirable baseline dependent multiplicative factors, such as bandpass shape and decorrelation, and
$f_0$ is the central frequency. As before, the prime indicates that there is an apparent phase shift due to the
ionosphere. The error term $\col{{N}}_{jk,c,f_0}^{\,(\dM)}$ contains a rotated version of the noise, $\col{{N}}_{jk,f}$,
and residual sidelobes due to errors in the calibrator source subtraction.

%----------------------------------------------------------------------------------------------------------------------%

\subsection{Ionospheric Refraction Measurements}
\label{Ionospheric Refraction Measurements}

The averaged and peeled visibilities given in (\ref{eqn: rotated vis}) contain the four instrumental polarization
products for calibrator $c$, with some added noise. These are the data that will be used to fit an ionospheric phase
ramp, as described in \cite{DocumentId=???}. Since the visibilities are dominated by flux from a single direction, they
can be converted to Stokes parameters, and only the Stokes I visibilities will be considered. Flux densities of
calibrator sources are expected to vary smoothly with frequency, and these variations will be catalogued. For the
purposes of combining phase data from across the entire observing band, they are divided out so that
$|I_{jk,c,f_0}^{\prime(\dM)}|$ can be replaced with $I_{c}$. The instrument Jones matrices will also vary with
frequency, but they are expected to vary in time on scales of minutes, and are assumed to be constant on the ionospheric
phase time scales of 10s of seconds. To deal with these variations, equations (\ref{eqn: response}) and (\ref{eqn:
rotated vis}) show that an estimate, $\mat{\I}_{jk,c,f}^{\prime(\dM)}$, of the sky visibility matrix for source $c$, can
be made using the visibilities calibrated with a recent gain solution

\begin{equation}
\label{eqn: calibrated response}
\begin{array}{llcl}
&
\mat{\I}_{jk,c,f_0}^{\prime(\dM)}
  & \approx & \mat{J}_{j,c,f_0}^{-1}\,\col{V}_{jk,c,f_0}^{\prime(\dM)}\,\mat{J}_{k,c,f_0}^{\dagger\,-1} \\
 &&& \\
 && \approx & \displaystyle\mat{\I}_{c,f_0}^{}\exp\{-i\delta\phi_{jk,c,f_0}\} \\
 &&& \\
\Rightarrow & I_{jk,c,f_0}^{\prime(\dM)}
  & \approx & \displaystyle I_{c}\exp\{-i\delta\phi_{jk,c,f_0}\}. \\
\end{array}
\end{equation}

Since the visibilities are phased towards the source, the sky $l$ and $m$ coordinates for it should be zero, and there
should be no phase ramp. Suppose though that, as indicated in (\ref{eqn: rotated vis}) and (\ref{eqn: calibrated
response}), the ionosphere adds a relative phase shift that appears to move the calibrator slightly in the $l$ and $m$
directions, and that the offsets are $\alpha_c\lambda_0^2$ and $\beta_c\lambda_0^2$ respectively, where $\lambda_0$ is
the wavelength associated with frequency $f_0$. Equation (\ref{eqn: calibrated response}) becomes

\begin{equation}
\label{eqn: ionospheric approx}
\begin{array}{lcl}
I_{jk,c,f_0}^{\prime(\dM)} & \approx & I_{c}\,\exp\{-i2\pi(\alpha_cu_{jk,f_0}+\beta_cv_{jk,f_0})\lambda_0^2\} \\
&&\\
                           & \approx & I_{c} - i2\pi I_{c}(\alpha_cu_{jk,f_0}+\beta_cv_{jk,f_0})\lambda_0^2, \\
\end{array}
\end{equation}

\noindent where $u_{jk,f_0}$ and $v_{jk,f_0}$ are components of the baseline vector introduced in section
\ref{Introduction}, and the expansion only holds when $(\alpha_cu_{jk,f_0}+\beta_cv_{jk,f_0})\lambda_0^2\ll1$. Erickson
found the root-mean-square displacement of sources observed with the Clark Lake TPT telescope at 74 MHz to be about
1$^\prime$.1, \cite{Erickson1984}. Erickson only considered relatively long-period fluctuations ($\sim$1 hr), and we
anticipate the variations for 8-second periods to be much less than this, with root-mean-square displacements of several
arcseconds or less. For arcminute deviations, the expansion in (\ref{eqn: ionospheric approx}) breaks down on the long
MWA baselines, so only short baselines are used for the initial fits. We then track the short-period deviations.

Each visibility contains the sum of components from thousands of cosmic sources, with different strengths and random
phases, and this sum, along with additive thermal noise from the receiving system, is expected to be very close to
normally distributed. This does indeed appear to be the case for the simulated visibilities discussed section
\ref{Examples}, with and without source peeling. If the noise is independent with variance $\sigma_{jk,f_0}^{2}$, and
$\Re()$ and $\Im()$ are the real and imaginary operators respectively, the least squares solutions for $I_{jk,c}$,
$\alpha_c$ and $\beta_c$ in (\ref{eqn: ionospheric approx}) are

\begin{equation}
\label{eqn: bob's solution}
\begin{array}{lcl}
I_{c}    & = & \displaystyle
               \frac{\sum_{jkf}\Re\left(I_{jk,c,f}^{\prime(\dM)}\right)\sigma_{jk,f}^{-2}}
                    {\sum_{jkf}\sigma_{jk,f}^{-2}} \\
&&\\
\alpha_{c} & = & \left( a_{vv}A_u - a_{uv}A_v \right) / (2\pi I_{c}\Delta) \\
&&\\
\beta_{c}  & = & \left( a_{uu}A_v - a_{uv}A_u \right) / (2\pi I_{c}\Delta),\\
\end{array}
\end{equation}

\noindent where

\begin{equation}
\label{eqn: bob's sums}
\begin{array}{lcl}
a_{uv} & = & \displaystyle\sum_{jkf}u_{jk}v_{jk,f} \sigma_{jk,f}^{-2}\lambda_0^{4} \\
&&\\
A_{u}  & = & -\displaystyle\sum_{jkf}u_{jk}\Im\left(I_{jk,c,f}^{\prime(\dM)}\right) \sigma_{jk,f}^{-2}\lambda_0^{2} \\
&&\\
\Delta & = & a_{uu}a_{vv} - a_{uv}^2
\end{array}
\end{equation}

The sums on the right-hand side of (\ref{eqn: bob's solution}) can be partially generated in each compute node, and then
$\alpha_{c}$ and $\beta_{c}$ generated after the partial sums have been gathered together.

%----------------------------------------------------------------------------------------------------------------------%

\subsection{Instrumental Gain Measurements}
\label{Instrumental Gain Measurements}

If a sufficiently reliable ionospheric refraction measurement has been made, then the phase center of the averaged
visibility set can be moved to the new position, and (\ref{eqn: rotated vis}) can be corrected:

\begin{equation}
\label{eqn: re-rotated instrumental vis}
\begin{array}{lcl}
\col{V}_{jk,c,f_0}^{(\dM)}
& \approx & \displaystyle \col{V}_{jk,c,f_0}^{\prime(\dM)}\exp\{+i\delta\phi_{jk,c,f_0}\}\\
&&\\
& \approx & \displaystyle \col{\GP}_{jk,c,f_0} + \col{{N}}_{jk,c,f_0}^{\,(\dM)}.
\end{array}
\end{equation}

To estimate the gain of an antenna towards calibrator $c$, a similar approach to those described in
\cite{BhatnagarNityananda2001} and \cite{Rogstad2005} will be taken, where one uses all of the visibilities that were
measured with an antenna to constrain a simple model of its gain. However, a matrix form from \cite{Hamaker2000a} will
be followed here. This is a matrix least-squares problem in which one searches for the matrices $\mat{J}_{j,c,f_0}^{-1}$
that minimize

\begin{equation}
\label{eqn: voltage amplitude model 1}
\sum_{j=1}^{N_a}\sum_{k,k\neq j}^{N_a}
\left\|\mat{\I}_{jk,c,f_0}^{(\dM)} -
      \mat{J}_{j,c,f_0}^{-1}\,\widehat{\col{V}}_{jk,c,f_0}^{}\,\mat{J}_{k,c,f_0}^{\dagger\,-1}\right\|_F^2,
\end{equation}

\noindent where $N_a$ is the number of antenna tiles, $\|\mat{A}\|_F^2$ is the squared Frobenius norm of a matrix
$\mat{A}$, equal to the trace of $\mat{A}^{}\mat{A}^\dagger$, and $\widehat{\col{V}}_{jk,c,f_0}$ is a model of the
measured visibility matrix for source $c$. In early investigations solutions to (\ref{eqn: voltage amplitude model 1})
were unstable when there were significant sidelobes from other sources, and rearranging (\ref{eqn: voltage amplitude
model 1}) had more robust solutions:

\begin{equation}
\label{eqn: voltage amplitude model 2}
\sum_{j=1}^{N_a}\sum_{k,k\neq j}^{N_a}
\left\|\col{V}_{jk,c,f_0}^{(\dM)} -
      \mat{J}_{j,c,f_0}\,\widehat{\mat{\I}}_{c,f_0}^{}\,\mat{J}_{k,c,f_0}^\dagger\right\|_F^2.
\end{equation}

For each antenna (\ref{eqn: voltage amplitude model 2}) has the analytic solution

\begin{equation}
\label{eqn: jones matrix solution}
\begin{array}{rl}
\mat{J}_{j,c,f_0}
& = \displaystyle\left(\sum_{k,k\neq j}^{N_a}\col{V}_{jk,c,f_0}^{(\dM)}
                       \mat{J}_{k,c,f_0}\widehat{\mat{\I}}_{c,f_0}^\dagger\right) \times \\
&   \displaystyle\left(\sum_{k,k\neq j}^{N_a}\widehat{\mat{\I}}_{c,f_0}^{}\mat{J}_{k,c,f_0}^\dagger
                       \mat{J}_{k,c,f_0}\widehat{\mat{\I}}_{c,f_0}^\dagger \right)^{-1}. \\
\end{array}
\end{equation}

Equation (\ref{eqn: jones matrix solution}) can be used to estimate a new set of Jones matrices for each calibrator (or
to update matrices that are restricted to moving more slowly). Once the matrices have converged they can be left to
track slow changes in the gain for each calibrator.

Finally, it is worth pointing out that often, much of the information in the Jones matrices is known. For example, the
transformation from the sky polarization coordinates to instrumental polarization coordinates may be known to high
precision. If this is the case, it is straightforward to modify (\ref{eqn: voltage amplitude model 2}) so that the other
matrices absorb the known information, leaving only the unknown quantities for the fit.

%----------------------------------------------------------------------------------------------------------------------%

\subsection{Source Subtraction}
\label{Source Subtraction}

The dynamic range of power levels expected from the hundred or so strongest calibrators will be many orders of
magnitude. This is the driving factor behind the sequential approach discussed above; stronger sources are subtracted
before measurements are made for the weaker sources. A preliminary investigation along the lines of \cite{Condon1974}
indicates that the deepest we will be able to peel or clean is to a sidelobe noise floor of several hundred mJy. At this
point, all remaining point sources will be less than 5 sigma above the noise. This suggests that there should be on the
order of a few hundred calibrators available, if we peel as deeply as possible.

The subtraction step is straightforward, since all of the CML sources have had their apparent positions measured, as
described in section \ref{Ionospheric Refraction Measurements}, and have been calibrated using the antenna gain models
given in (\ref{eqn: jones matrix solution}). Before peeling, however, the models need to be multiplied by the complex
$\eta_{jk,c,f}$ factors. These were divided out when averaging to a central frequency in section \ref{Rotate
Visibilities and Sum over Frequency}.

It may be that subtracting 100\% of each source is not optimal, since noise in each measurement is added back into the
visibilities. Some modifications that work more like the CLEAN algorithm (see \cite{Hogbom1974}) are being considered,
and will be tested soon.

%----------------------------------------------------------------------------------------------------------------------%

\section{Discussion}
\label{Discussion}

\subsection*{Performance}

The relatively short cadence time of the RTS means we need to make compromises when designing real-time peeling process.
One of these compromises will be the number of sources peeled every 8 seconds. Several methods for calibrating antenna
gains are compared in \cite{Boonstra2003}. These are alternatives to the technique discussed in section
\ref{Instrumental Gain Measurements}. The number of complex multiplications required by most of the techniques scales as
$N_a^3$, where $N_a$ is the number of antennas (or, more generally, the number of receivers being correlated to form
visibilities). The technique that required the fewest computations was the logarithmic least squares (LOGLS) algorithm,
which scaled as $N_a^2$. The numbers given in \cite{Boonstra2003} for a single polarization version are $2N_a^2$
multiplications with an additional $16N_a^2$ for weighting.

How does this compare to the algorithm discussed in section \ref{Instrumental Gain Measurements}? In our application we
measure two polarizations, so if we consider each antenna as 2 polarized receptors the number of multiplications for the
LOGLS algorithm is $2(2N_a)^2$. There will probably be another factor of 2 since correlations between the receptors on
the same antenna will most likely need to be considered.\footnote{The authors of \cite{Boonstra2003} note that LOGLS is
not easily generalized to a dual-polarized telescope array, but, for the sake of comparison, suppose that such a
generalization exists.} The number of complex multiplications used to determine the calibration solutions in section
\ref{Instrumental Gain Measurements} is $O(24N_a^2)$, with an additional $O(12N_a^2)$ for weighting. This efficiency
appears to be about as high as one might reasonably expect to achieve.

We are not just dealing with a single source, however. The CML needs to pre-peel all of the calibrator sources, and
then, for each source, unpeel, rotate all of the visibilities, solve for the ionospheric offset, solve for the antenna
gains, and peel. Table \ref{tbl:FLOPS} shows the approximate number of floating-point operations used in each of these
steps. These numbers were obtained by listing the main operations in the inner loops of the routines and multiplying
each by our best estimate of the associated floating-point operations. These numbers are not exact; they are provided as
a rough indication of where processing time will be spent. Also, the number of sources processed by each of the routines
need not be the same. For example, we might peel and make gain measurements for 50 sources, but make ionospheric
measurements on a few hundred more (which requires only the third and fourth rows).

\begin{table}[htb]
  \centering
  \caption{Approximate floating-point operations required for each source in the CML (with 512 antenna tiles and a
           single frequency channel).}
  \begin{tabular}{ c c }
    {\bf Routine} &
    {\bf Floating-point operations (millions)} \\
  \hline
    peeling (applied 3 times)       &  31 \\
    rotate and accumulate           &  26 \\
    ionospheric sums and rerotation &  21.5 \\
    measure tile gains              &  37 \\
  \hline
    total                           & $O$(180) \\
  \end{tabular}
  \label{tbl:FLOPS}
\end{table}

While the algorithm scales well, there is still on the order of 180 million floating-point operations required during
each 8 second calibration cycle. For the whole array, some of the rows in the table need to be multiplied by the number
of frequency channels (768), while others need to be multiplied by the number of frequency sub-bands ($\sim50$). They
also need to be multiplied by the number of sources, as discussed in the previous paragraph. We anticipate a few
trillion floating-point operations for the CML over the 8 seconds. This does not include various overheads such as
memory access that will increase the number of operations by a factor of a few.

\subsection*{Options for Reducing the Load}

Each ionospheric refraction measurement can use data from every baseline, polarization and frequency. That is over 1.5
gigasamples for a single 2D offset. For sources that are only used for ionospheric measurements, all of the processing
listed in table \ref{tbl:FLOPS} will be reduced if we only use a subset of the visibilities. Most of the baselines are
short, and many of those will be redundant. Furthermore, long baselines do not see as much of the extended galactic
structure as short baselines, and they measure the apparent offset with higher angular resolution. Initial
investigations suggest that we may be able to ignore more than 99\% of the short baselines for the stronger sources.

Since the tile gains are changing slowly, we do not necessarily need to make measurements for every source every 8
seconds. This can be exploited by making measurements for the strongest sources every 8 seconds, and cycling through
subsets of the other sources. This way we still have gain measurements distributed across the sky every few minutes when
we make fits for the tile beams, but we only run the full peel algorithm on 20 or 30 sources at a time.

Rather than solving for direction-dependent parameters sequentially, one could fit for all of the calibrator sources
simultaneously. In fact, the peeling algorithm is really just a robust and efficient method for reducing the number of
unknowns and finding the multivariate solutions. One can also reduce the degrees of freedom by changing the fit
parameters to quantities that do not change (or change slowly) with time or frequency, and solve using multiple
snapshots, as discussed in detail in \cite{vanderTol2007}. For the MWA, we hope to be able to use slowly varying
direction-dependent dipole gains and phases to describe the primary beam of each tile. Once we have high quality
measurements of our tile beams in the field, we will look more closely at fitting these dipole parameters directly.
However, source subtraction will then be based on these dipole fits, not on direct tile gain measurements.

\subsection*{Foreground Subtraction}

One of the primary challenges in the search for a signature from the EoR is that of foreground subtraction. At best the
signal will be several orders of magnitude weaker than the galactic foreground and it is important that we understand
the nature of the residuals from the calibration and peeling process. One of the drawbacks of peeling strong sources in
real time and then averaging the resulting images together is that any residuals are also averaged into the mix. All of
the calibration data will be stored in a database for use in offline processing, and peeling errors can be assessed and
reduced at that stage. However, if there is any concern that residuals from the peeling process might mimic the EoR
signal then peeling can be used for calibration only, and images formed from unpeeled visibilities.

There is significant effort going into techniques for removing foregrounds during offline processing. These include
techniques that exploit spectral differences in the foregrounds and the EoR signature (see for example
\cite{Morales2006} and the discussion and references in \cite{Furlanetto2006}), and the rotation measure synthesis
techniques discussed in \cite{Burn1966} \& \cite{Brentjens2005}. Direction-dependent deconvolution techniques such as
the one described in \cite{Bhatnagar2008} will also be essential for imaging.

\subsection*{Radio Frequency Interference}

Of concern for any telescope operating at MWA frequencies is radio frequency interference (RFI). Even though the
Murchison site is extremely radio quiet \cite{Bowman2007a}, the array will still have to deal with some RFI. This
includes communication and military satellite signals, reflections of FM radio broadcasts, and natural interference such
as lightning. Due to the low spectral occupancy of the RFI, the high quality polyphase filter banks used to isolate
frequencies, and the campaign-mode operation of the array, we will adopt the traditional strategy of flagging and
ignoring contaminated data before imaging. Missing frequency or time samples can be accounted for in the weighting of
the various least-squares algorithms.

%----------------------------------------------------------------------------------------------------------------------%

\section{Examples}
\label{Examples}

To test the CML, we generate simulated visibilities using MAPS, the MIT Array Performance Simulator \cite{Wayth2008}.
Briefly, MAPS models all physical processes of a radio interferometer from the ionosphere, through the analog
beamforming in the array tiles to time and frequency averaging in a correlator. MAPS uses polarized receptors and a
polarized model sky to generate model visibilities in linear or circular polarization products. The model sky includes a
large-scale diffuse component, based on \cite{deOliveira-Costa} with additional polarized flux, plus a catalog of
southern point sources based on \cite{Wright1990} and \cite{Mauch2003}.

The tile beams for MWA simulations consist of 16 dual polarization receptors in a 4x4 array. Receptors can have
non-equal complex gains, which allows us to include realistic differences in the tile primary beams that we might expect
due to the analog parts of the system. In the examples that follow, complex Gaussian noise was added to the gain of each
receptor.

MWA's synthesized beam has modest resolution, which allows us to use a very realistic full-sky model as an input to MAPS
by simultaneously including large-scale diffuse structure and thousands of point sources. As such, the simulations show
a large range of correlated power depending on baseline length. MAPS also implements a model ionosphere to change the
relative path length for each baseline in each look direction. The model uses the International Reference Ionosphere
\cite{Rawer1978} for large scale structure, and a Kolmogorov spectrum to add turbulence at smaller scales. The turbulent
power was a set fraction of the total power -- about 2\% -- and was repeated over the sky in patches of about 1 square
degree. Phase variations due to traveling ionospheric disturbances are not included in these simulations, but they have
recently been added to MAPS. Finally, 200K thermal noise was added to the visibilities. In this paper, while the
description is fully polarized and there is significant instrumental polarization, we only use unpolarized input since
testing of the MAPS software with polarized input is incomplete. Polarized sources will be needed to fully describe the
polarized response of each antenna and the Faraday rotation state of the ionosphere \cite{Sault-etal.1996},
\cite{Hamaker2000a}.

\begin{figure*}[ht]
\begin{center}
$\begin{array}{ccc}
\mbox{\small{a) 0 sources peeled}} &
\mbox{\small{b) 10 sources peeled}} &
\mbox{\small{c) 100 sources peeled}} \\
\includegraphics[width=5.8cm]{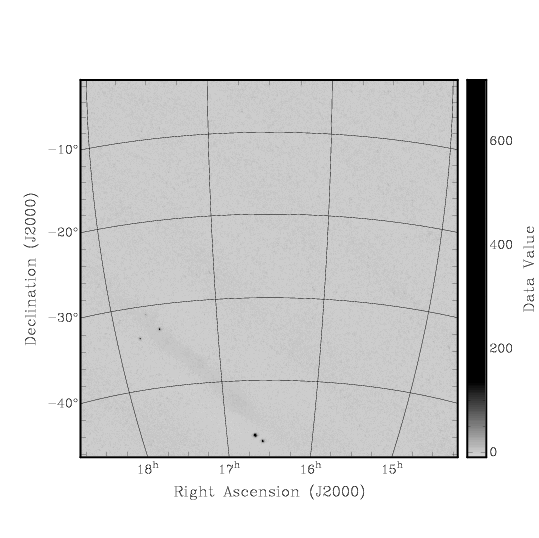} &  
\includegraphics[width=5.8cm]{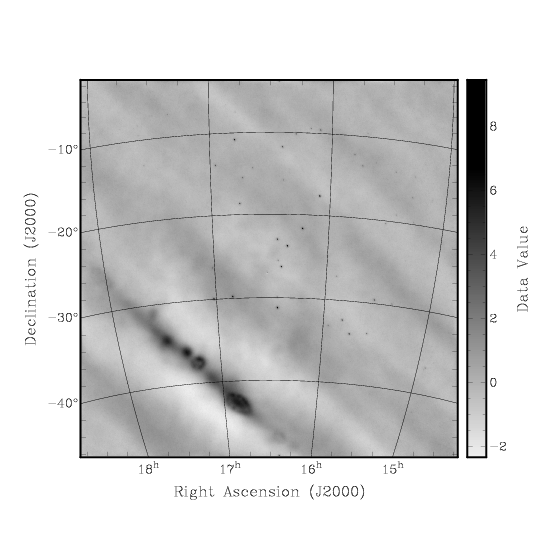} &
\includegraphics[width=5.8cm]{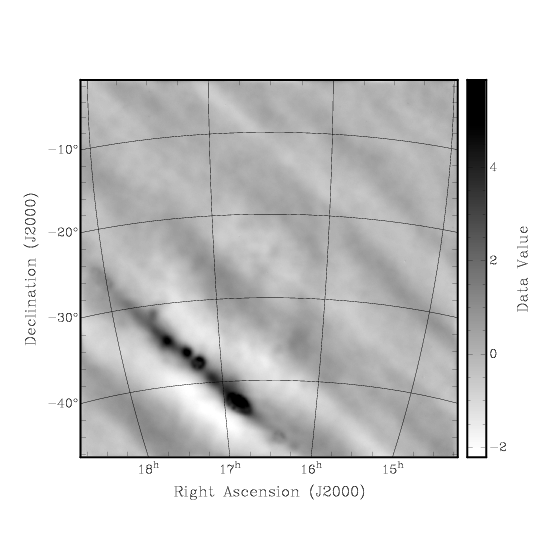} \\
\end{array}$
\end{center}
\caption{Uncalibrated Stokes I images made after peeling. Shown are $45^\circ\times45^\circ$ images, however sources are
         peeled from anywhere in the sky, depending on their apparent strength. As more sources are peeled, weaker point
         sources are revealed until, in this field, the galactic center and its sidelobes dominate. Gray-scales are set
         by the minimum and maximum pixel values, which have units of Jy/beam.}
\label{plot:images}
\end{figure*}

MAPS was used to generate a series of visibility sets at a local sidereal time at the MWA site of 16.5 hours, with the
antenna beams pointed at the zenith. Fig. \ref{plot:images} shows three images created after the CML had converged. The
only difference in the processing of each image is the number of sources that were peeled. Fig. \ref{plot:images}a shows
the case for no peeling, where a few strong point sources dominate the image. There is also a slight hint of diffuse
galactic structure in the lower left. As background radio sources are peeled, weaker sources and the diffuse galactic
foreground become apparent. In Fig. \ref{plot:images}c all of the stronger sources have been analyzed and peeled, and
the galactic center and its wavy sidelobes completely dominate.

\begin{figure}[ht]
\centering
\includegraphics[width=9cm]{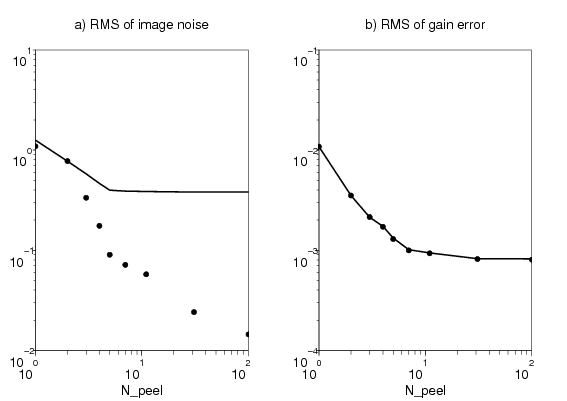}\\
\caption{Convergence as a function of the number of source peeled: a) RMS of the image noise (Jy/beam), b) RMS of the
         gain error for the strongest source. The solid curves show full simulations that include point sources,
         galactic emission and ionospheric effects, while the points represent simulations that only had point sources.
         The similarity of the gain error curves suggests that the algorithm performs well in the face of an ionosphere
         and galactic emission.}
\label{plot:Convergence with Npeel}
\end{figure}

Fig. \ref{plot:Convergence with Npeel}a shows the image noise RMS as a function of the number of sources being peeled
(in one of the instrument polarizations). The solid curve is from the simulations described above, the points are for
images generated using the point sources only (no galactic emission or ionosphere). After the first few strong sources
are peeled away, the image RMS becomes dominated by the galaxy. To reduce the image RMS beyond this point, more
sophisticated foreground subtraction algorithms need to be employed, such as those discussed in section \ref{Discussion}
(one should note that EoR observations will not be made in a field that contains the galactic center). Fig.
\ref{plot:Convergence with Npeel}b shows the RMS of the tile gain error towards the strongest calibrator as a function
of the number of sources being peeled. Here the curve and the points converge at the same rate, indicating that for this
source the algorithm is not limited by the ionosphere or the galactic emission. In these examples, we have weighted down
visibilities from short baselines, since they see all of the galactic emission (Fourier components of angular features
larger than the reciprocal of the baseline length cancel destructively, so we give more weight to the longer baselines).

\begin{figure}[ht]
  \centering
  \includegraphics[width=9cm]{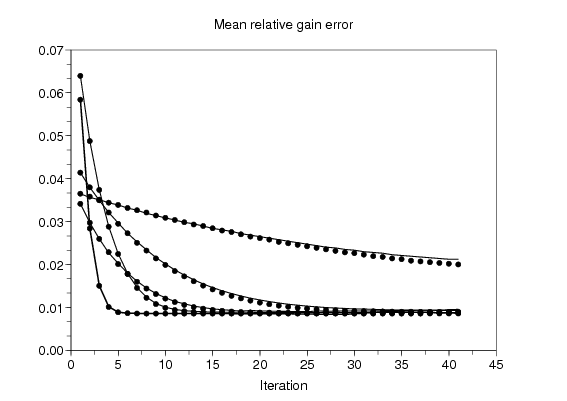}\\
  \caption{Convergence as a function of time. Shown is the mean relative gain error for the 5 strongest sources.
           The solid curves represents full simulations that include point sources, galactic emission and ionospheric
           effects, and the points show simulations that only had point sources.
           Again, the convergence is not being limited by the ionosphere or the galactic emission.}
  \label{plot:Convergence with Niter}
\end{figure}

Also of importance is how quickly convergence is achieved. At the end of each iteration (for the MWA, each successive
iteration is associated with a new set of visibilities, i.e., they occur 8 seconds apart), the new solutions are used to
update the old solutions. Since the data are noisy, a weighted average of new and old solutions are used to set the
weights for the next iteration. In these simulations, the weights used for each source were based on the estimated
contribution of the source to the visibilities. In other words, strong sources are updated quickly (the strongest being
allowed to update by 50\% with each iteration), while weak sources are adapted more slowly, some by only a few percent
each time. Fig. \ref{plot:Convergence with Niter} shows the mean gain errors for the 5 strongest sources, normalized by
the gain towards each source. The errors are reduced until convergence is limited by a local minimum in the minimization
process. The relative errors appear to converge to the same level, which has not been investigated.

The properties and limits of the convergence shown in Fig. \ref{plot:Convergence with Niter} is of the utmost
importance. In the current simulations, typically a few tens of calibrators converge well, but some weaker sources can
be significantly affected by sidelobes and converge to local minima of (\ref{eqn: voltage amplitude model 2}). A
quantitative analysis of the situation is currently underway, and several options are being tested. These include
investigations of the calibration bandwidth (sidelobes of distant sources will decorrelate more as the bandwidth of the
sub-bands is increased), large pre-peels (approximate subtraction of extra catalogue sources helps reduce the size of
local minima), and incorporating all-sky ionospheric and primary beam models to improve peels for weak sources. The
primary beam fits are seen to work better when the calibrators are not just ranked by their apparent flux density, but
also their position in the beams, so that a larger fraction of them lie in the structured antenna sidelobes.

%----------------------------------------------------------------------------------------------------------------------%

\section{Summary}
\label{Summary}

We have described a general approach for making measurements of strong point sources that can be used in the calibration
process of wide-field, low-frequency radio arrays. This approach has been adopted for the MWA, and there is an ongoing
effort to develop the required software, as well as to understand the benefits and limitations of the approach.

We have used simulated visibility data to show that the peeling algorithm works well in situations that are of major
concern for future radio telescopes: crowded fields, strong galactic emission, and ionospheric refraction. The algorithm
exhibits fast convergence, which is important since sources will be moving in and out of antenna sidelobes and the
algorithm needs to be able to keep up with the antenna gain and phase changes, as well as changes in the ionosphere.
Critical parts of the process are shown to be computationally efficient, and parts of the system lend themselves to
significant levels of optimization.

The next step is a detailed analysis of the convergence properties of the algorithms, and a series of tolerance tests to
investigate how the algorithms will behave in the various conditions we expect to encounter. This includes observations
of weak emission that is masked by significant polarized diffuse foregrounds, high dynamic range observations close to
the sun, and observations in the presence of severe ionospheric conditions, such as during activation and recombination
of the ionosphere. As data from the initial deployment of antennas become available in late 2008 and early 2009, we will
get a clearer picture of how harmful phenomena such as source variability (due to ionospheric scintillation, for
instance) and dipole mutual coupling (which will affect our tile beam models) can be, and these can be worked into the
tolerance tests.

%----------------------------------------------------------------------------------------------------------------------%

\section*{Acknowledgments}

We would like to acknowledge the international MWA collaboration for its continued input and support of this work.

%----------------------------------------------------------------------------------------------------------------------%

%----------------------------------------------------------------------------------------------------------------------%

\end{document}